 \documentclass[smallabstract,smallcaptions]{dccpaper}

\usepackage{epsfig}
\usepackage{amsmath}
\usepackage{amssymb}
\usepackage{color}
\usepackage{url}
\usepackage{multirow}

\newlength{\figurewidth}
\newlength{\smallfigurewidth}

\newcommand{\RN}[1]{%
  \textup{\uppercase\expandafter{\romannumeral#1}}%
}

\begin{document}

\title
{\large
\textbf{Deep Learning-based Image Compression with \\ Trellis Coded Quantization}
}

\author{%
Binglin Li$^{\ast}$, Mohammad Akbari$^{\ast}$, Jie Liang$^{\ast}$, and Yang Wang$^{\dag}$\\[0.5em]
{\small\begin{minipage}{\linewidth}\begin{center}
\begin{tabular}{ccc}
$^{\ast}$School of Engineering Science & \hspace*{0.5in} & $^{\dag}$Department of Computer Science\\
Simon Fraser University && University of Manitoba \\
Burnaby, BC, Canada && Winnipeg, MB, Canada\\
\url{{binglinl, akbari, jiel}@sfu.ca} && \url{ywang@cs.umanitoba.ca}
\end{tabular}
\end{center}\end{minipage}}
}

\maketitle
\thispagestyle{empty}

\begin{abstract}
Recently many works attempt to develop image compression models based on deep learning architectures, where the uniform scalar quantizer (SQ) is commonly applied to the feature maps between the encoder and decoder. In this paper, we propose to incorporate trellis coded quantizer (TCQ) into a deep learning based image compression framework. A soft-to-hard strategy is applied to allow for back propagation during training. We develop a simple image compression model that consists of  three subnetworks (encoder, decoder and entropy estimation), and optimize all of the components in an end-to-end manner. We experiment on two high resolution image datasets and both show that our model can achieve superior performance at low bit rates. We also show the comparisons between TCQ and SQ based on our proposed baseline model and demonstrate the advantage of TCQ. 
\end{abstract}

\section{Introduction}
%Conventional compression standards such as JPEG~\cite{wallace1992jpeg} and JPEG2000~\cite{jpeg2000} mainly include the image transformation, quantization, entropy coding for the encoder and these steps are optimized separately. The decoder is composed of the inverse operations to reconstruct the original image. 
The goal of designing the optimal image codec is to minimize the distortion $ D $ between the original image and the reconstructed image subject to the constraint of the bitrate $ R $. As the entropy $ H $ is the lower bound of bitrate $ R $, the optimization can be formulated as minimizing $ D + \lambda H $, where $ \lambda>0 $ is the tradeoff factor. 
Recently many works~\cite{balle2016end, theis2017lossy, mentzer2018conditional} attempt to develop image compression models based on deep learning architectures. In their approaches, a uniform scalar quantizer (SQ) is commonly applied to the feature maps between the encoder and decoder.
%In information theory, a quantization step represents a sequence of data $ X^{n}=\lbrace X_{1},X_{2}, \cdots, X_{n} \rbrace $ by an index $ fn(X)^{n} $ in $ {1,2,\cdots, 2^{2nR}} $, where $ R $ is the number of bits used to encode each sample. When $ n=1 $, it quantizes each sample individually, which we call scalar quantizer (SQ). Vector quantization (VQ) is a generalization of SQ where $ n>1 $ to quantize a group of samples jointly. When $ n $ increases, VQ is more close to the Gaussian R-D function. However, a larger $ n $ leads to more complex search computation as well as memory consumption.
As the codewords are distributed in a cubic and the corresponding Voronoi regions induced by SQ are always cubic, SQ cannot achieve the R-D bound~\cite{jpeg2000}. 
Vector quantization (VQ) has the optimal performance, but the complexity is usually high. 
%Trellis coded quantizer (TCQ) is a structural scalar quantizer, but it can achieve similar performance to VQ with lower complexity~\cite{jpeg2000, marcellin1990trellis}. 
Trellis coded quantizer (TCQ) is a structured VQ, and it can achieve better performance than SQ with modest computational complexity~\cite{marcellin1990trellis}.
It is shown in~\cite{marcellin1990trellis} that for memoryless uniform sources, a 4 state TCQ can achieve 0.87dB higher SNR than SQ for 4 bit$ / $sample. 

In this paper, motivated by the superior performance of TCQ over SQ in traditional image coding, we propose to use TCQ to replace the commonly used SQ in a deep learning based image compression model. The soft-to-hard strategy~\cite{agustsson2017soft} is applied to allow for back propagation during training. 
%Although in deep learning models, the performance improvement may be restricted by the powerful learning ability of neural network for compensattion, it is still worthwhile to 
To the best of our knowledge, we are the first to investigate the performance of TCQ in a deep learning based image compression framework.
Our implementation allows for batch processing amenable to the mini-batch training in deep learning models, which greatly reduces the training time. 

The entropy coding can further reduce the bitrate without impacting the reconstruction performance. One way to apply it in deep learning model is to use offline entropy coding method during testing~\cite{akbari2019dsslic}. This method is not optimized for the bitrate as the network is not explicitly designed to minimize the entropy. In this paper, we adopt the PixelCNN++~\cite{salimans2017pixelcnn++} to model the probability density function on an image $ x $ over pixels from all channels as $ p(x)=\prod_{i}p(x_{i}|x_{<i}) $, where the conditional probability only depends on the pixels above and to the left of the pixel in the image. A cross entropy loss is followed to estimate the entropy of the quantized representation to jointly minimize the R-D function. 

%Similar to previous works, we train an individual model for each bit rate to obtain the R-D curve.
%In~\cite{balle2016end, theis2017lossy} they use continuous functions for approximation to compuate the entropy during training, and optimize it along with the distortion errors.
%Previous works train an individual model for each bit rate. Similar to~\cite{akbari2019dsslic}, we use residual coding to obtain the performance for later bit rate, which consists of the second layer of the entire coding scheme. Different from~\cite{toderici2015variable} that LSTM recurrent networks are used to further encode the residuals, we use one of the conventional codecs for the residual coding because of its efficiency. 

Our contributions are summarized as follows. We propose to incorporate TCQ into a deep learning based image compression framework. The image compression framework consists of encoder, decoder and entropy estimation subnetworks. They are optimized in an end-to-end manner. We experiment on two commonly used datasets and both show that our model can achieve superior performance at low bit rates. We also compare TCQ and SQ based on the same baseline model and demonstrate the advantage of TCQ. 

\begin{figure*}
\begin{center}
%\fbox{\rule{0pt}{2in} \rule{0.9\linewidth}{0pt}}
   \includegraphics[width=0.8\linewidth]{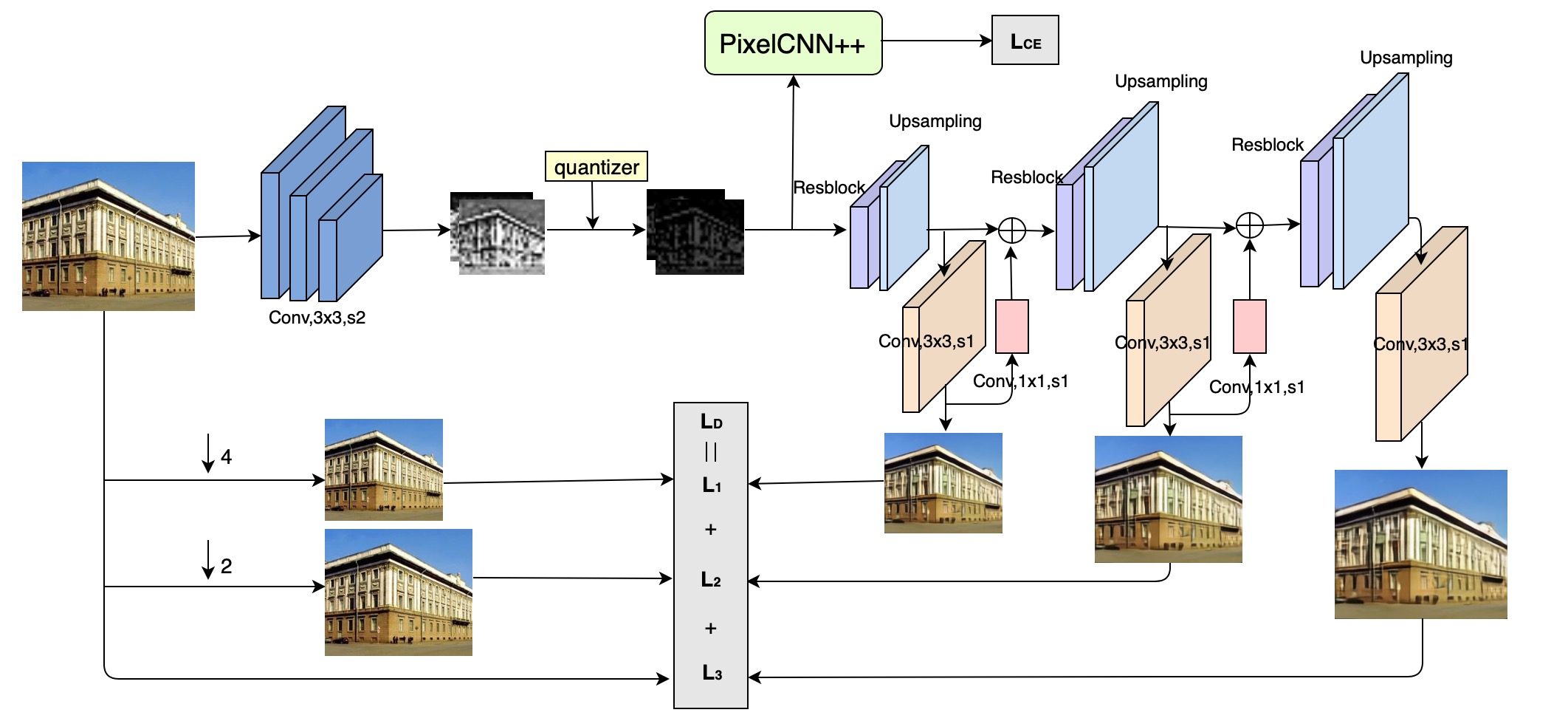}
\end{center}
   \caption{Overview of our proposed deep image compression model. The encoder has three consecutive strided convolutional layers to reduce the input size from 256$ \times $256 to 32$ \times $32. After the TCQ is applied, the quantized feature representations are used as input to two branches. One goes to the decoder network to generate image from resolution 64$ \times $64 to 128$ \times $128, and then to 256$ \times $256. These three losses are added together to be the distortion loss $ L_{D} $. The other one goes to the entropy model (pixelCNN++) to produce the probability matrix of pixels based on previous pixels optimized by the cross entropy loss $ L_{CE} $. ``Conv, $a \times a$, s\textit{b}" denotes convolutional layer with $ a \times a $ kernels and a stride of \textit{b}. ``Resblock" is from~\cite{he2016deep} but without BatchNorm layers. ``$ \downarrow $\textit{c} " represents downsampling by a factor of \textit{c}.}
\label{fig:overview}
%\label{fig:onecol}
\end{figure*}
\section{Related Work}
There has been a line of research on deep learning based image compression, especially autoencoders with a bottleneck to learn compact representations. The encoder maps the image data to the latent space with reduced dimensionality, and the decoder reconstructs the original image from the latent representation. 
\subsection{Quantization in DNN}
%As the entropy coding is lossless and the quantization can cause some degree of information loss during encoding, a good quantizer is crucial to reconstruction performance. 
Several approximation approaches have been proposed to allow the network to back-propagate through the quantizer during training. 
In~\cite{toderici2015variable, li2018learning}, a binarization layer is designed in the forward pass and the gradients are defined based on a proxy of the binarizer. 
Ball\'{e} \textit{et. al.}~\cite{balle2016end} stochastically round the given values by adding noise and use the new continuous function to compute the gradients during the backward pass. 
Theis \textit{et. al.}~\cite{theis2017lossy} extend the binarizer in~\cite{toderici2015variable} to integers and use straight-through estimator in the backward pass. 
In~\cite{agustsson2017soft}, a soft quantization in both forward and backward passes is proposed. The model needs to learn the centers and change from soft quantization to hard assignments during training by an annealing strategy. 
In~\cite{mentzer2018conditional}, the authors apply the nearest neighbors to obtain fixed centers, and the soft quantization in~\cite{agustsson2017soft} is used during the backward pass.
%A non-uniform quantizer is proposed in~\cite{cai2018deep} and the codebooks are learnt by the Lloyd-Max algorithm. The quantizer and the auto-encoder model are learnt in an alternative way.
%These work attempt to approximate the gradients with 1 so that the output value can unchangably go back to update the output of previous layer, or change the quantization process as an approximate continuous function to make it differentiable. 
%There are also some new quantizers proposed in DNN compression works~\cite{jung2019learning}, where they truncate the ranges of the activations and weights to balance the memotry requirement and the classificaiton accuracy. 

\subsection{Image Compression based on DNN}
%For lossy image compression, JPEG~\cite{wallace1992jpeg} has been widely used in digital photographs. Later several more sophisticated standards were developed such as JPEG 2000~\cite{jpeg2000}, WebP~\cite{webp}, and Better Portable Graphics (BPG)~\cite{bpg}. 
%In these approaches, image compression is achieved by nonlinearly transforming image pixels to a quantized latent representation, which is in the form of bitstreams, and then followed with losslessly compressing these latents by entropy coding.
With the quantizer being differentiable, in order to jointly minimize the bitrate and distortion, we also need to make the entropy  differentiable.
For example, in~\cite{balle2016end, theis2017lossy}, the quantizer is added with uniform noise. The density function of this relaxed formulation is continuous and can be used as an approximation of the entropy of the quantized values. 
In~\cite{agustsson2017soft}, similar to the soft quantization strategy, a soft entropy is designed by summing up the partial assignments to each center instead of counting.
In~\cite{mentzer2018conditional, li2018learning}, an entropy coding scheme is trained to learn the dependencies among the symbols in the latent representation by using a context model. These methods allow jointly optimizing the R-D function.

\section{Proposed Approach}
Our model follows the encoder-decoder framework. Different from the previous works that apply a uniform scalar quantizer (SQ) after the encoder network, we propose to use trellis coded quantizer (TCQ) to enhance the reconstruction performance. The whole framework is trained jointly with our entropy model.

\subsection{Encoder and Decoder}
Since our goal is to study the gain of TCQ and SQ, we only use a simple encoding and decoding framework.
Our encoder network consists of three layers of convolutional layers with a stride of 2 to downsample the input. Each convolutional layer is followed by a ReLU layer. We remove BatchNorm~\cite{ioffe2017batch} layers as we find removing them gives us better reconstruction performance. We add one more convolutional layer to reduce the channel dimension to a small value \textit{e.g.} 8 to get a condensed feature representation $ F $. A $ Tanh $ layer is followed to project $ F $ to continuous values between -1 and 1. 
Then a quantizer is applied to quantize the feature maps to discrete values.
For the decoder network, we use PixelShuffle~\cite{shi2016real} layer for upsampling. Inspired by~\cite{xu2018attngan}, we adopt two intermediate losses after each upsampling operation to force the network to generate images from low resolution to high resolution progressively as shown in Fig. \ref{fig:overview}.  

\subsection{Trellis Coded Quantizer}
%\begin{figure}[t]
%\begin{center}
%\fbox{\rule{0pt}{2in} \rule{0.9\linewidth}{0pt}}
%   \includegraphics[width=0.7\linewidth]{fig/tcq_structure.jpg}
%\end{center}
%   \caption{An example of 4-state trellis structure.}
%\label{fig:tcq_diagram}
%\label{fig:onecol}
%\end{figure}
\begin{figure}[t]
\begin{minipage}[b]{0.45\linewidth}
   %\centering
    \includegraphics[width=0.9\linewidth, height=30mm]{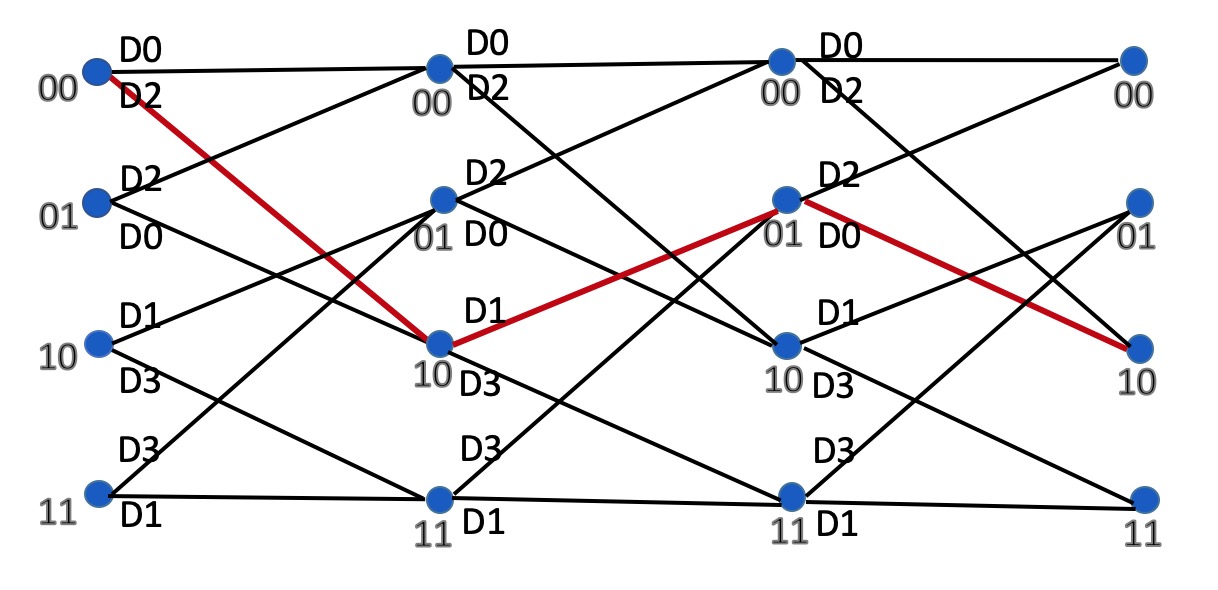}
    \caption{An example of 4 state trellis structure.}
    \label{fig:tcq_diagram}
\end{minipage}%
%    \hfill%
\hspace{\fill}
\begin{minipage}[b]{0.45\linewidth}
   \centering
	\begin{tabular}{ccc}
	%\fbox{\rule{0pt}{2in} \rule{0.9\linewidth}{0pt}}
   \includegraphics[width=0.18\linewidth, height=7mm]{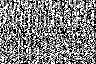}&
   \includegraphics[width=0.18\linewidth, height=7mm]{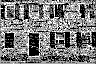} &
   \includegraphics[width=0.18\linewidth, height=7mm]{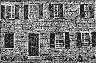}\\
    \includegraphics[width=0.18\linewidth, height=16mm]{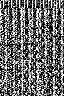}&
   \includegraphics[width=0.18\linewidth, height=16mm]{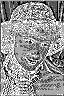} &
   \includegraphics[width=0.18\linewidth, height=16mm]{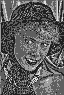}\\
   {\small (a)} & {\small (b)} & {\small (c) }
\end{tabular}
    \caption{(a) indexing method \RN{1} for TCQ, (b) indexing method \RN{2} for TCQ, (c) SQ.}
    \label{fig:tcq_indices}
\end{minipage} 
\end{figure}

\textbf{Forward Pass:} Trellis coded quantizer (TCQ) is applied in JPEG2000~\cite{jpeg2000} part \RN{2}. Different from JPEG2000 where the input for TCQ is fixed given an image block, when embedded in deep neural networks, the input for TCQ is updated in each iteration during training.
The forward pass for TCQ is similar to the original implementation in~\cite{jpeg2000}. %, where we use Viterbi algorithim~\cite{forney1973viterbi} to save the intermediate results. 
In essence, TCQ aims to find a path with minimum distortion from the start symbol to the last symbol based on the particular diagram structure. Figure~\ref{fig:tcq_diagram} shows a trellis structure with 4 states. 
For $ R $ bit/symbol, a quantizer with $ 2^{R+1} $ quantization levels is created. These $ 2^{R+1} $ reconstruction points can be obtained by a uniform quantizer. As the last layer of our encoder is a $ Tanh $ function, we have $ V_{max}=1 $ and $ V_{min}=-1 $. The quantization step is $ \Delta=\dfrac{2}{2^{R+1}} $. A reconstruction point $ c_{j} $ ($ j=1, 2, \cdots, 2^{R+1} $) is obtained by $ c_{j}=-1 + \Delta/2 + (j-1 )\times \Delta$.
Next all the reconstruction levels are partitioned into four subsets $ {D_{0}, D_{1}, D_{2}, D_{3}} $ from left to right to form four sub-quantizers.
%From left to right, we repeatly assign each level to $ {D_{0}, D_{1}, D_{2}, D_{3}} $. For $ R $ bit/symbol, each subquantizer has $ 2^{R-1} $ codewords and needs $ R-1 $ bits to represent. Therefore, TCQ actually has $ 2^{R+1} $ levels in total distributed for 4 subquantizers compared to a scalar quantizer (SQ) with $ R $ bit/symbol.
%For one edge $ w_{i\rightarrow i}^{j\rightarrow j+1} $, based on the TCQ diagram in Fig.~\ref{fig:tcq_diagram}, only one subquantizer is chosen for this edge. We compute the quantization error within the codewords in this subquantizer and keep the smallest value as the weight for this edge. As mentioned above, the edge weight is cached and only the state transition that gives the minimum result for current point will be kept for later usage.
Then different subsets are assigned to different branches of the trellis, so that different paths of the trellis can try different combinations to encode an input sequence. Each node only needs to record the input branch that has the smallest cost.
After obtaining the minimum distortion for the last symbol, we trace back to get the optimal path as shown in red in Fig.~\ref{fig:tcq_diagram} for instance. With this optimal path, 1 bit $ q $ is used to indicate which branch to move for next symbol, and the last $ R-1 $ bits $ {b_{1}b_{2}\cdots b_{R-1}} $ are used to indicate the index of codeword from the corresponding sub-quantizer. Here we call it indexing method \RN{1} .
%Note that there are also only two edges from one point to next one. 
%In Fig.~\ref{fig:tcq_diagram}, the branch number for the three symbols are 1, 0, 1 respectively (0 for first branch and 1 for second branch). 

\textbf{Backward Pass:} In order to make a quantizer differentiable, the most common way is to use straight-through estimator~\cite{bengio2013estimating} where the derivative of the quantizer is set to 1. However, we find that such backward method tends to converge slowly for TCQ. As the TCQ changes the distribution of the input data, this inconsistency may make it hard for the network to update weights in the right direction. Similar to~\cite{mentzer2018conditional}, given reconstruction points $ C=\lbrace c_{1}, c_{2}, \cdots, c_{L}\rbrace $ ($ L=2^{R+1} $), we use the differentiable soft quantization during the backward pass.
\begin{equation}
\tilde{Q}(z) = \Sigma_{j=1}^{L}\dfrac{exp(-\sigma||z-c_{j}||)}{\Sigma_{l=1}^{L}exp(-\sigma||z-c_{l}||)}c_{j}
\end{equation}
where $ \sigma $ is a hyperparameter to adjust the ``softness" of the quantization.

\textbf{Discussions:} One issue for the TCQ implementation is that the time and memory complexity are both proportional to the number of symbols. Previous implementation usually flattens the input block into a sequence. Because pixels in one feature map are more correlated than pixels in other feature maps, we consider each feature map as an input for TCQ. For feature maps with size $ B\times C\times H\times W $ ($ B $ is the batch size for the network, $ C $ is the number of channels, $ H $ and $ W $ are the height and width), we reshape the size as $ BC \times HW $, where $ BC $ is the batch size for TCQ and $ HW $ is the number of symbols in a feature map, which reduces the processing time. 

The other issue is that the conventional indexing method \RN{1} mentioned above %(one bit $ q $ to indicate the branch number and the last $ R-1 $ bits $ {b_{1}b_{2}\cdots b_{R-1}} $ to represent the index for the codeword in a subquantizer) 
brings in randomness for the indices of a feature map as shown in Fig.~\ref{fig:tcq_indices} (a).
%n Fig.~\ref{fig:tcq_indices} (a) we consider to put $ q $ as the last bit \textit{i.e.} $ {b_{1}b_{2}\cdots b_{R-1}}q $ to alleviate the randomness. 
The reason is that the branch bit $ q $ depends on the optimal path in trellis structure and it does not carry any relationship among each symbol. From JPEG2000~\cite{jpeg2000}, we have two union-quantizers $ A0={D0 \bigcup D2} $ and $ A1={D1 \bigcup D3} $. As pointed in~\cite{marcellin1994entropy}, given a node in the diagram, the codeword that can be chosen is either from $ A0 $ or $ A1 $. Therefore, because of the particular structure of the trellis, all $ R $ bits can be used to represent the indices for the union-quantizer $ A0 $ and the same applies to $ A1 $. For example, in Fig. \ref{fig:tcq_diagram}, assume we receive the initial state $ 01 $ during decoding. Only $ D0 $ or $ D2 $ sub-quantizer will be chosen for this symbol. As the indices for $ D0 $ and $ D2 $ are all different, we get the corresponding unique codeword based on the received $ R $ bits. Then we easily know which sub-quantizer ($ D0 $ or $ D2 $) is chosen and accordingly the branch number. We call it indexing method \RN{2}. Fig.~\ref{fig:tcq_indices} (b) gives the indices of a feature map resulting from the indexing method \RN{2}. 

%\begin{figure}[t]
%\begin{center}
%\begin{tabular}{ccc}
%\fbox{\rule{0pt}{2in} \rule{0.9\linewidth}{0pt}}
%   \includegraphics[width=0.2\linewidth, height=20mm]{fig/img_00000_c_00_tcq_branch.png}&
%   \includegraphics[width=0.2\linewidth, height=20mm]{fig/img_00000_c_00_tcq.png} &
%   \includegraphics[width=0.2\linewidth, height=20mm]{fig/img_00000_c_00_scalar.png}\\
%    \includegraphics[width=0.2\linewidth, height=40mm]{fig/img_00003_c_00_tcq_branch.png}&
%   \includegraphics[width=0.2\linewidth, height=40mm]{fig/img_00003_c_00_tcq.png} &
%   \includegraphics[width=0.2\linewidth, height=40mm]{fig/img_00003_c_00_scalar.png}\\
%   {\small (a)} & {\small (b)} & {\small (c)}
%\end{tabular}
%\end{center}
%   \caption{(a) First column are from indexing with branch number for TCQ. (b) Second column are from indexing with new method for TCQ. (c) Third column are from indexing for SQ.}
%\label{fig:tcq_indices}
%\label{fig:onecol}
%\end{figure}

\subsection{Entropy Coding Model}
%For the above model, we can compute the bit rate directly by $ r = \dfrac{C\times H\times W\times R}{I_{h}\times I_{w}} $, where $ C\times H \times W $ is size of feature maps for one image, $ R $ is the number of bit for each symbol, $ I_{h} $ and $ I_{w} $ are image height and width. 
%However, the forementioned autoencoder model is not optimized for entropy coding. This equation can be seen as nominal (pre-entropy coding) bit rate. 
The aforementioned autoencoder model is not optimized for entropy coding. We can model the conditional probability distribution of a symbol based on its context~\cite{mentzer2018conditional}. The context should be only related to previous decoded symbols, and not use the later unseen symbols. We employ PixelCNN++~\cite{salimans2017pixelcnn++} model for the entropy coding model. 
%Compared to PixelCNN~\cite{van2016conditional}, PixelCNN++ improves to model the joint distribution of each pixel and only needs $ H\cdot W $ times forward passes during decoding compared to $ 3\cdot H \cdot W $ times in PixelCNN. 
We replace the last layer of PixelCNN++ model in implementation\footnote{https://github.com/pclucas14/pixel-cnn-pp} with a softmax function 
%to directly output the probability of current symbol given the context from previous symbols. 
so that a cross entropy loss can be used during training. This loss is viewed as an estimation of entropy for the quantized latent representation.
Assume we have $ R $ bits to encode each symbol and a $  C\times H \times W $ dimensional feature map $ F $, the PixelCNN++ model outputs a $ 2^{R}\times C\times H \times W  $ probability matrix. Encoding is done row by row and each row orders from left to right. With the probability matrix, we encoder the indices of the feature maps by Adaptive Arithmetic Coding (AAC)\footnote{https://github.com/nayuki/Reference-arithmetic-coding} to get the compressed representation. 
During decoding, for the first forward pass, we input the pre-trained PixelCNN++ model with a tensor with all zeros. This first forward pass gives distributions for entries $ p(z_{c=1:C, i=1, j=1}) $ where $ (c, i, j) $ is a position in the feature map $ F $. Then we decode the indices along the channel dimension by AAC. Based on the received initial states, we recover the symbols at $ ({\small c=1:C, i=1, j=1}) $. The following decoding steps are based on the conditional probability 
\begin{equation}
p(z_{c=1:C, i=u, j=v}|Context) = 
\text{PixelCNN++}(T_{{\small \lbrace 1:C, 1:u, 1:(v-1) \rbrace \bigcup \lbrace 1:C, 1:(u-1), v:W \rbrace }})_{c,i,j}
\end{equation}
where $ T_{{\small 1:x, 1:y, 1:z}} $ is a tensor with decoded symbol at location $ \lbrace (c,i,j) | 1\leq c\leq x, 1\leq i\leq y, 1\leq j\leq z\rbrace $ and zeros otherwise. When $ u=1 $, $ {\small \lbrace 1:C, 1:(u-1), v:W \rbrace}=\O $. When $ v=1 $, $ {\small \lbrace 1:C, 1:u, 1:(v-1) \rbrace}=\O $. As the decoding proceeds, the remaining zeros will be replaced by the decoded symbols progressively. 
\section{Experiment}
\subsection{Dataset}
We use ADE20K dataset~\cite{zhou2017scene} for training and validation. We test on Kodak PhotoCD image dataset\footnote{http://r0k.us/graphics/kodak/} and Tecnick SAMPLING dataset~\cite{asuni2014testimages}. ADE20K dataset contains 20K training and 2K validation images. Kodak PhotoCD image dataset and Tecnick SAMPLING dataset include 24 512$ \times $768 images and 100 1200$ \times $1200 images respectively.

\subsection{Training Details}
We crop each input image by 256$ \times $256 during training and test on the whole images. During training, we use a learning rate of 0.0001 at the beginning, and decrease it by a factor 0.4 at epoch 80, 100 and 120. Training is stopped at 140 epochs and we use the model that gives the best validation result for testing. We set the batch size as 18 and run the training on one 12G GTX TITAN GPU with the Adam optimizer.
We use 4 quantization levels and increase the channel size from $ \lbrace 4, 6, 8, 12, 16 \rbrace $ to control the bitrate.
Compression performance is evaluated with Multi-Scale Structural Similarity (MS-SSIM) by bits per pixel (bpp) and we use MS-SSIM loss in Eq.~\ref{Eq:MSSSIM} during training.
\begin{equation}
L_{\text{MS-SSIM}} = 100(1-L_{D}(\text{MS-SSIM}(\tilde{X},X))) + \lambda L_{CE}
\label{Eq:MSSSIM}
\end{equation}
The first term is the distortion error and the second term is the cross entropy loss for pixelCNN++ model. $ \lambda $ is a hyperparameter and set to 1.

\subsection{Baselines}
We compare our results with conventional codecs and recent deep learning based compression models. JPEG~\cite{wallace1992jpeg} results are obtained from ImageMagick\footnote{https://imagemagick.org}. JPEG2000 results are from MATLAB implementation and BPG results are based on 4:2:0 chroma format\footnote{http://bellard.org/bpg}. For deep learning based image compression models, we either collect from the released test results or plot the rate-distortion curves from the published papers.

\begin{figure}[t]
\begin{center}
\begin{tabular}{cc}
%\fbox{\rule{0pt}{2in} \rule{0.9\linewidth}{0pt}}
   \includegraphics[width=0.47\linewidth]{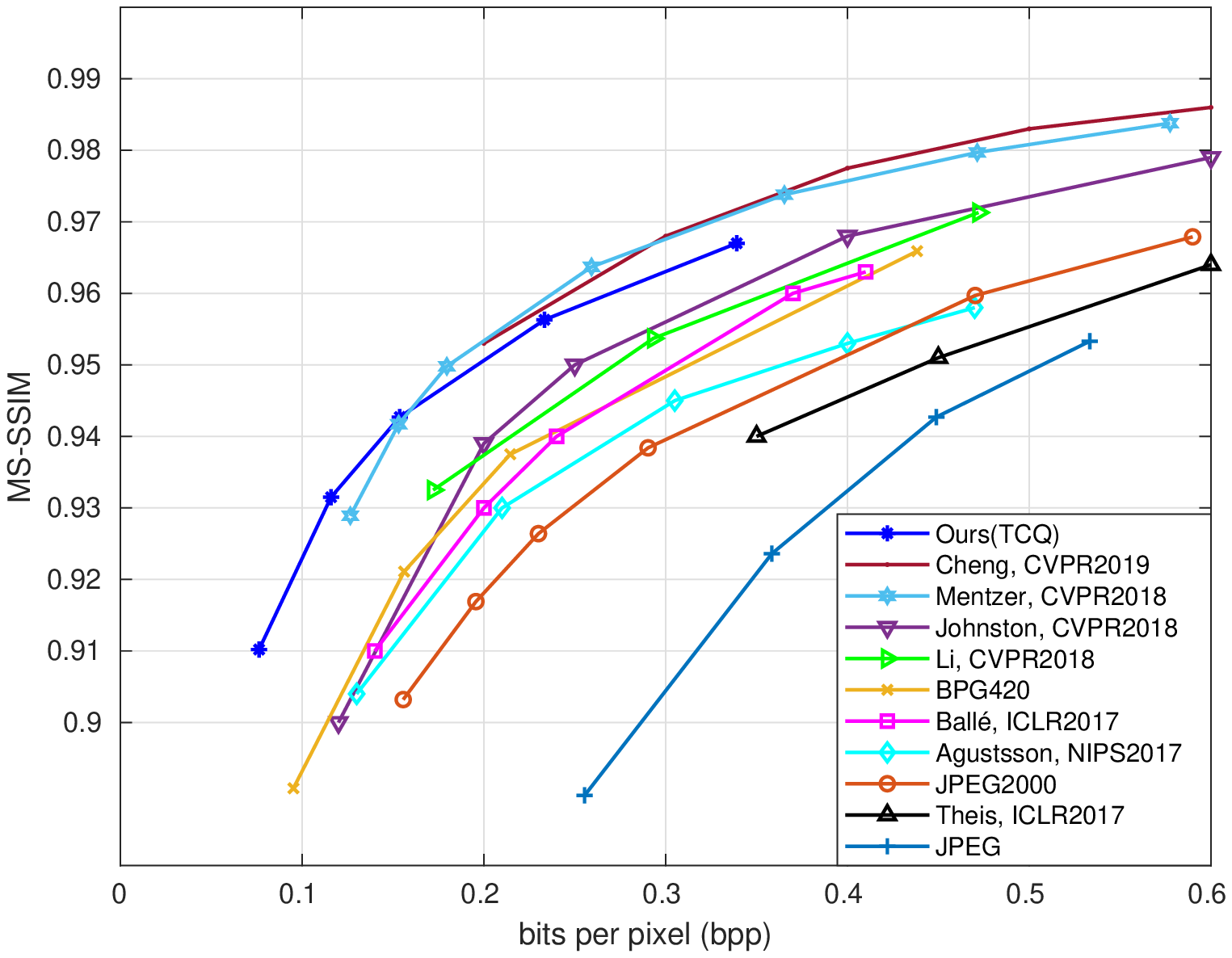} & 
   \includegraphics[width=0.47\linewidth]{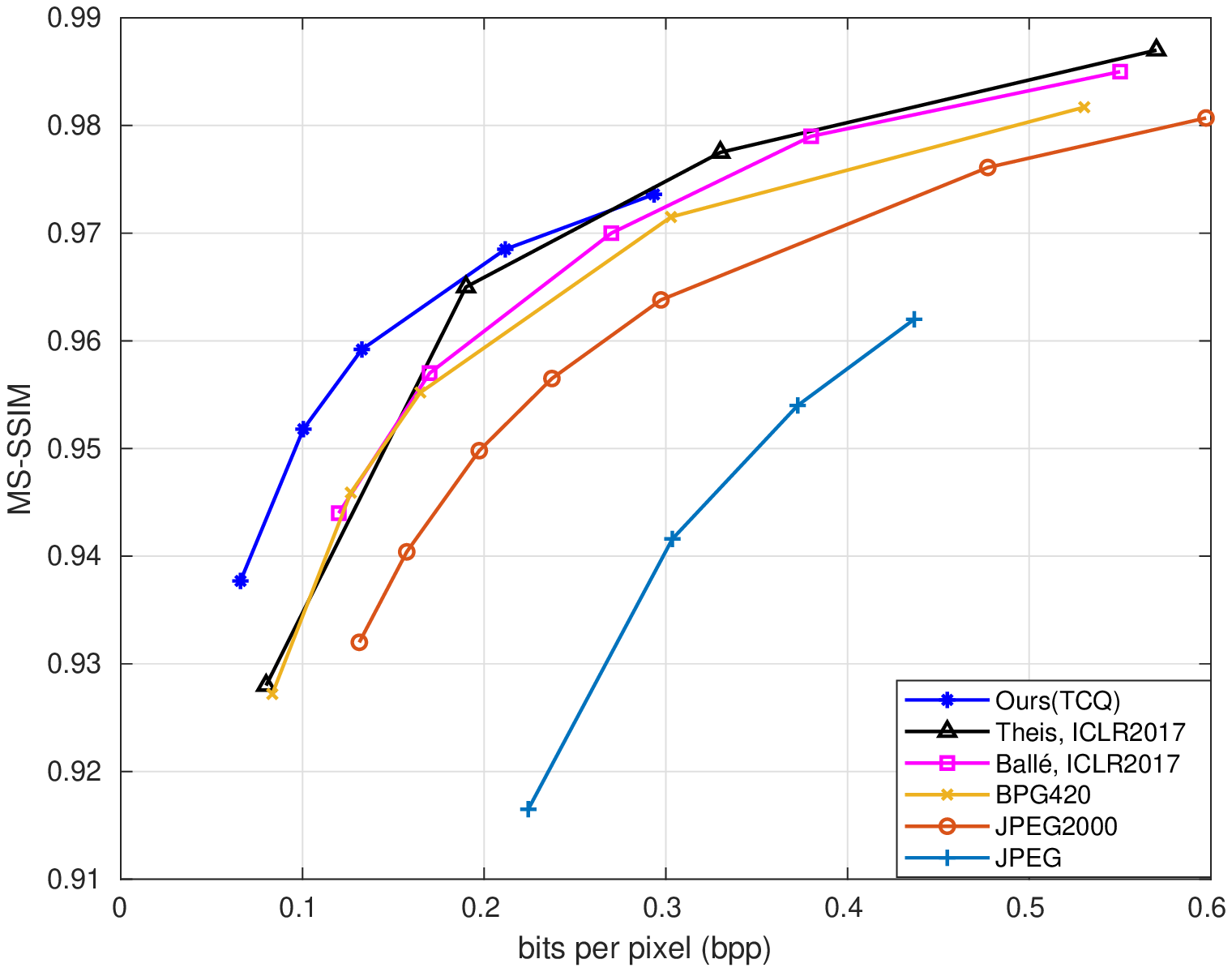} \\
   {\small (a)}  & {\small (b) }
\end{tabular}
\end{center}
   \caption{(a) MS-SSIM/bpp on Kodak dataset. (b) MS-SSIM/bpp on Tecnick dataset.}
\label{fig:results}
\end{figure}

\begin{table}[t]
\begin{center}
\caption{\label{tab:tcq_scalar_msssim}%
Performance comparisons between TCQ and SQ using MS-SSIM loss for training}
\begin{tabular}{|c|c|c|c|c|}
\hline
\multirow{2}{*}{quantizer} & \multicolumn{2}{c|}{Kodak dataset} & \multicolumn{2}{c|}{Tecnick dataset} \\
\cline{2-5}
 &  PSNR(dB)/bpp & MS-SSIM/bpp & PSNR(dB)/bpp & MS-SSIM/bpp \\
\hline
SQ    & 	     24.54/0.077		&    0.9028/0.077		    &       26.14/0.068 		&  0.9326/0.068	\\
TCQ  & 	  	24.95/0.076		&	  0.9102/0.076		    &		26.82/0.066     &  0.9377/0.066    \\
\hline
SQ    & 	    25.66/0.117		&    0.9259/0.117		    &       27.63/0.104 		&  0.9493/0.104     	\\
TCQ  & 	  	25.85/0.116		&	  0.9315/0.116		    &		27.86/0.101		&  0.9518/0.101    		\\
\hline
SQ    & 	     26.23/0.157		&    0.9386/0.157 		    &       28.32/0.139		&  0.9572/0.139		\\
TCQ  & 	  	26.47/0.154		&	  0.9427/0.154		    &		28.45/0.133 		&  0.9592/0.133    	\\
%\hline
%SQ    & 	     27.05/0.232		&    0.9532/0.232		    &       29.51/0.208		&  0.9674/0.208		\\
%TCQ  & 	  	27.28/0.233		&	  0.9563/0.233		    &		29.56/0.212		&  0.9685/0.212   		\\
\hline
\end{tabular}
\end{center}
\end{table}

\begin{table}[!p]
\begin{center}
\caption{\label{tab:tcq_scalar_mse}%
Performance comparisons between TCQ and SQ using MSE loss for training}
\begin{tabular}{|c|c|c|c|c|}
\hline
\multirow{2}{*}{quantizer} & \multicolumn{2}{c|}{Kodak dataset} & \multicolumn{2}{c|}{Tecnick dataset} \\
\cline{2-5}
 &  PSNR(dB)/bpp & MS-SSIM/bpp & PSNR(dB)/bpp & MS-SSIM/bpp \\
\hline
SQ    & 	   24.86/0.064			&    0.8715/0.064    	&       25.94/0.054 		&  	 0.9091/0.054	\\
TCQ  & 	   25.23/0.062			&	  0.8824/0.062   	&		 26.66/0.052		     &       0.9189/0.052   \\
\hline
SQ    & 	   25.81/0.098			&    0.8992/0.098	     &       27.35/0.081		     &  	 0.9308/0.081	\\
TCQ  & 	   26.35/0.096			&	  0.9090/0.096 	&		 27.96/0.078 		&       0.9373/0.078  \\
\hline
SQ    & 	   26.56/0.133			&    0.9178/0.133    	&       28.18/0.112			&  	0.9427/0.112 	\\
TCQ  & 	   26.89/0.130			&	  0.9232/0.130		&		 28.65/0.110 			&       0.9473/0.110  \\
\hline
\end{tabular}
\end{center}
\end{table}

\begin{figure}[!p]
\begin{center}
\begin{tabular}{cccc}
%\fbox{\rule{0pt}{2in} \rule{0.9\linewidth}{0pt}}
 \includegraphics[width=0.20\linewidth, height=20mm]{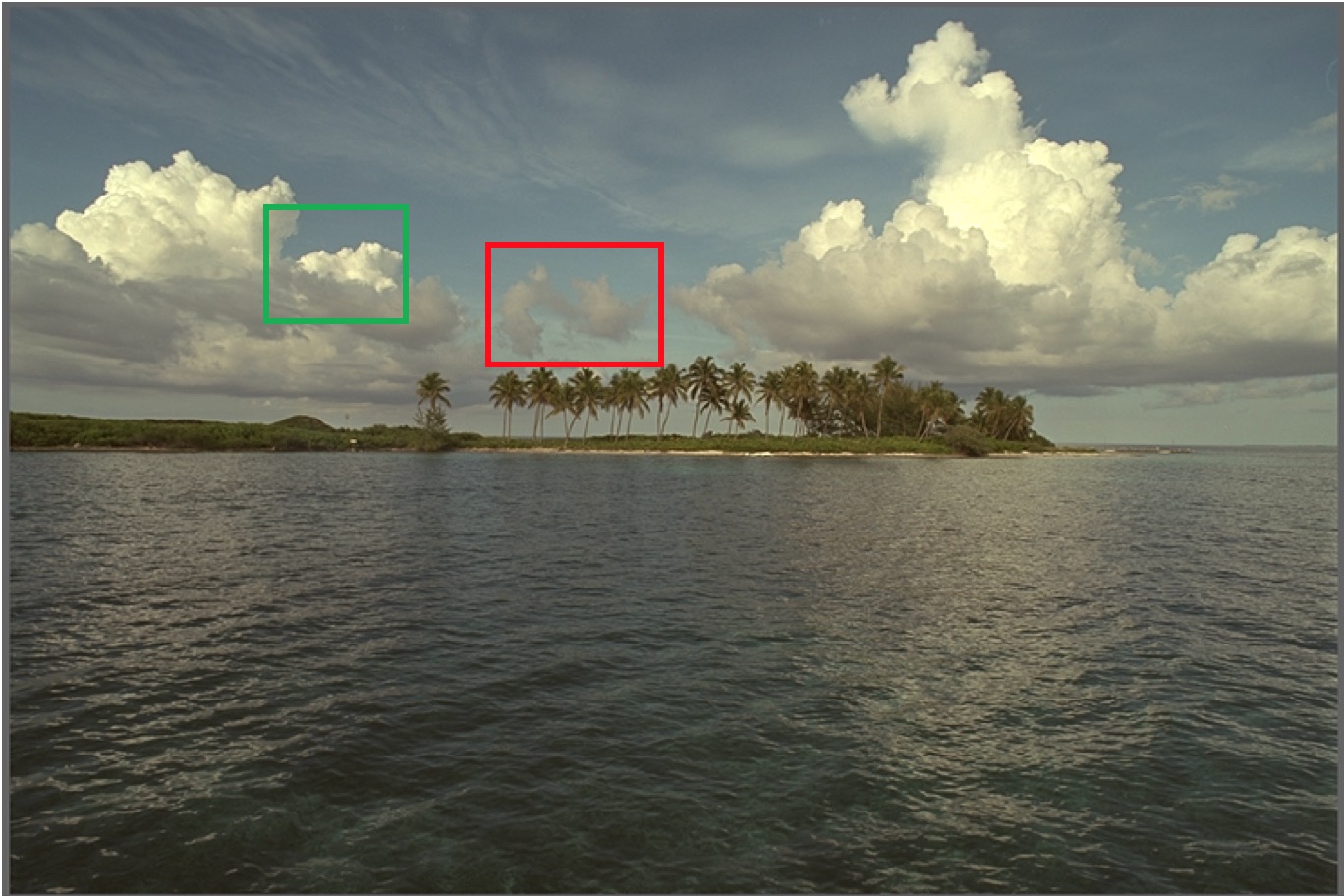} &
  \includegraphics[width=0.20\linewidth, height=20mm]{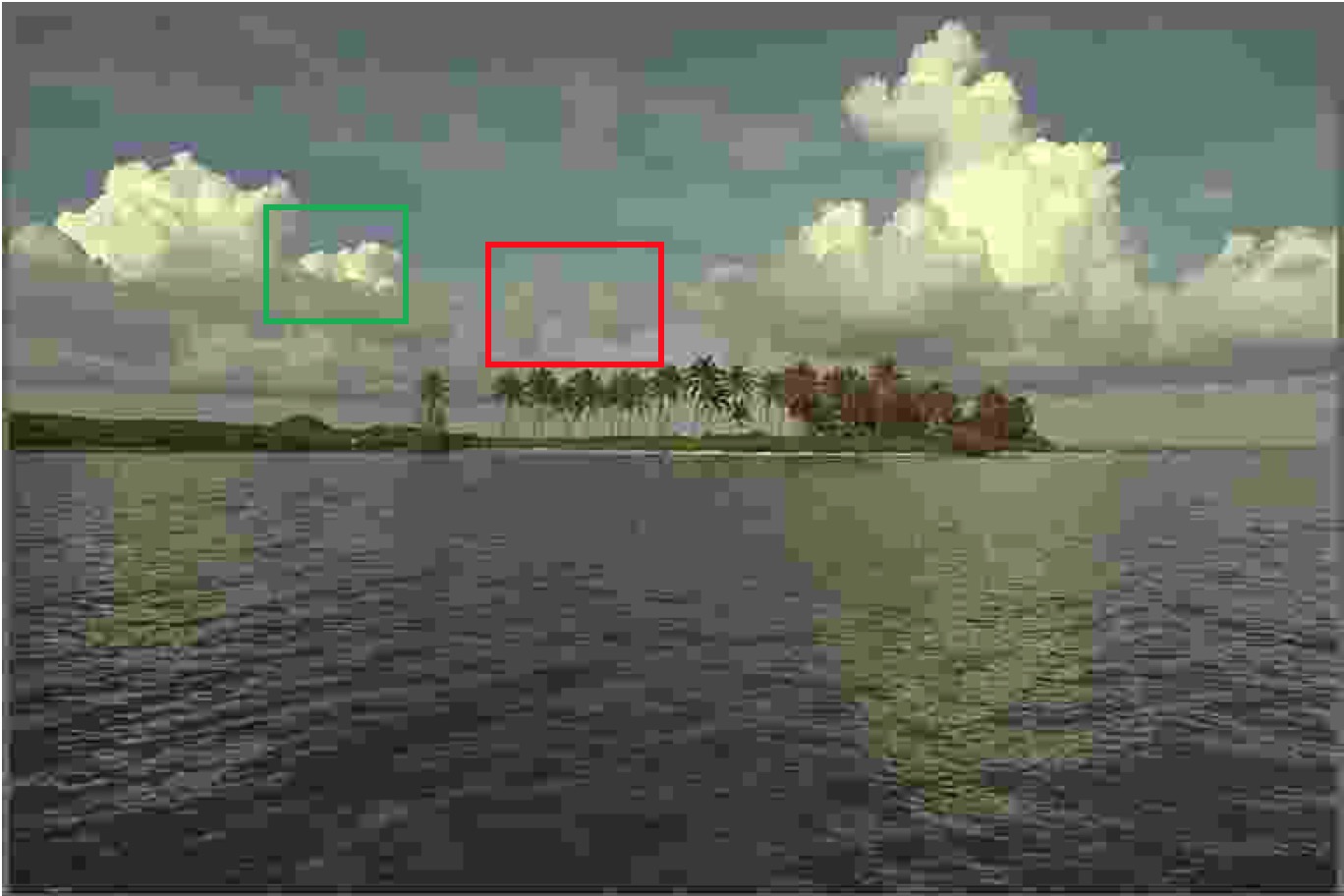}&
   \includegraphics[width=0.20\linewidth, height=20mm]{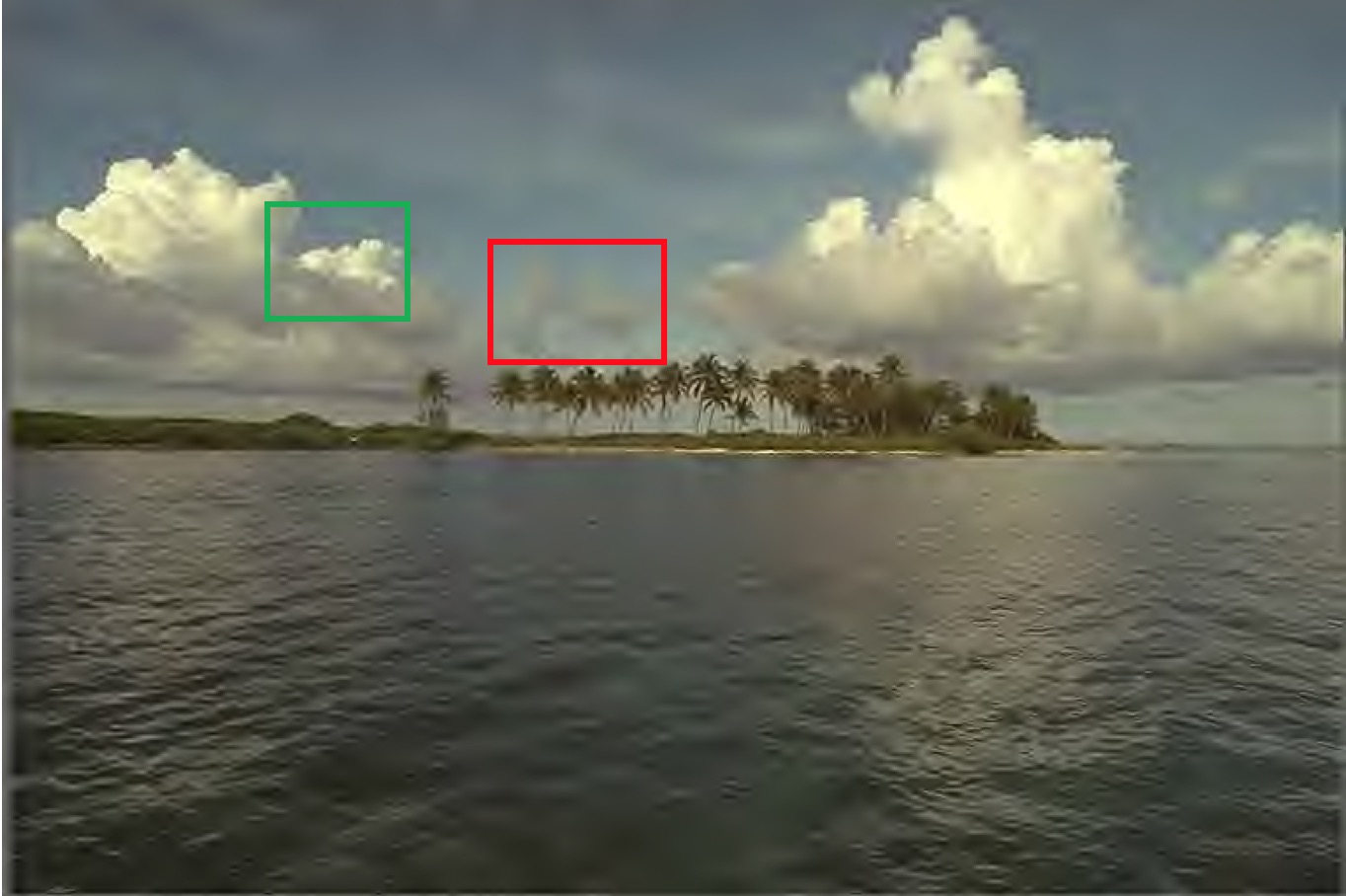}&
   \includegraphics[width=0.20\linewidth, height=20mm]{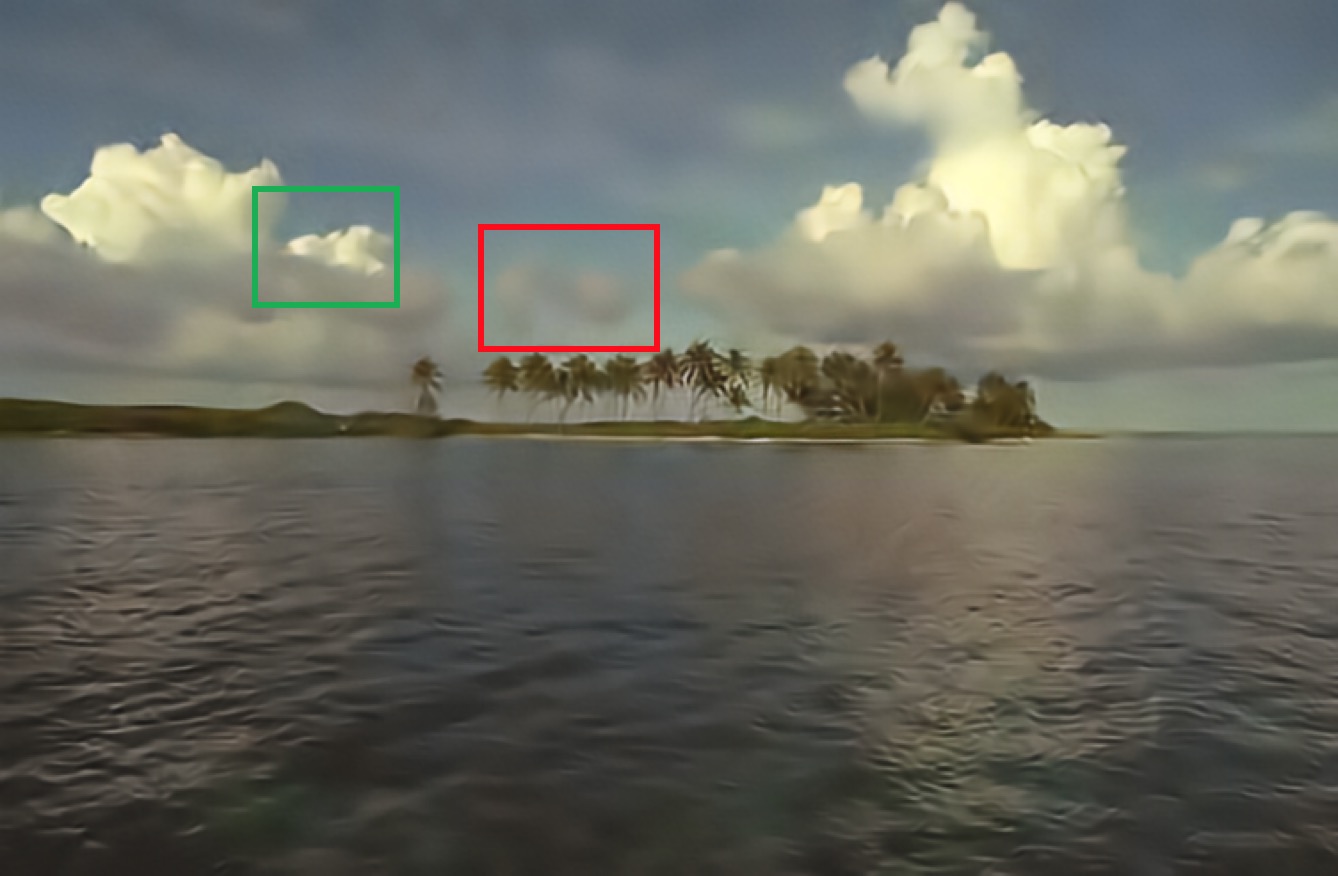} \\
\fboxsep=0mm
\fboxrule=1.2pt   
 \fcolorbox{green}{black}{\includegraphics[width=0.20\linewidth, height=20mm]{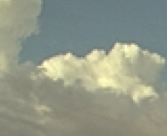}} &
 \fboxsep=0mm
\fboxrule=1.2pt 
 \fcolorbox{green}{black}{\includegraphics[width=0.20\linewidth, height=20mm]{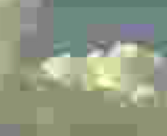}}&
 \fboxsep=0mm
\fboxrule=1.2pt 
 \fcolorbox{green}{black}{\includegraphics[width=0.20\linewidth, height=20mm]{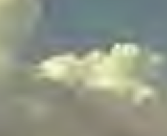}}&
 \fboxsep=0mm
\fboxrule=1.2pt 
 \fcolorbox{green}{black}{\includegraphics[width=0.20\linewidth, height=20mm]{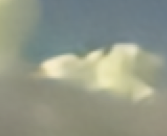}} \\  
 \fboxsep=0mm
\fboxrule=1.2pt 
  \fcolorbox{red}{black}{\includegraphics[width=0.20\linewidth, height=17mm]{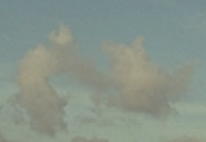}} &
  \fboxsep=0mm
\fboxrule=1.2pt 
  \fcolorbox{red}{black}{\includegraphics[width=0.20\linewidth, height=17mm]{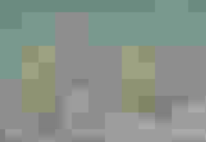}}&
  \fboxsep=0mm
\fboxrule=1.2pt 
  \fcolorbox{red}{black}{\includegraphics[width=0.20\linewidth, height=17mm]{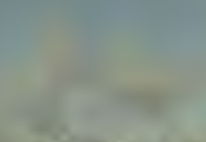}}&
  \fboxsep=0mm
\fboxrule=1.2pt 
  \fcolorbox{red}{black}{\includegraphics[width=0.20\linewidth, height=17mm]{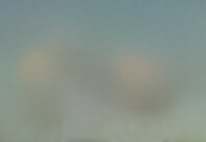}} \\
    									  & {\small 0.8177/0.128} 		& {\small 0.9028/0.118} 	&  {\small 0.9139/0.117}  \\   
        {\small (a) Original}  & {\small (b) JPEG}  			& {\small (c) JPEG2000}  	& {\small (d) Ball\'{e}~\cite{balle2016end}}  \\
   &
   \includegraphics[width=0.20\linewidth, height=20mm]{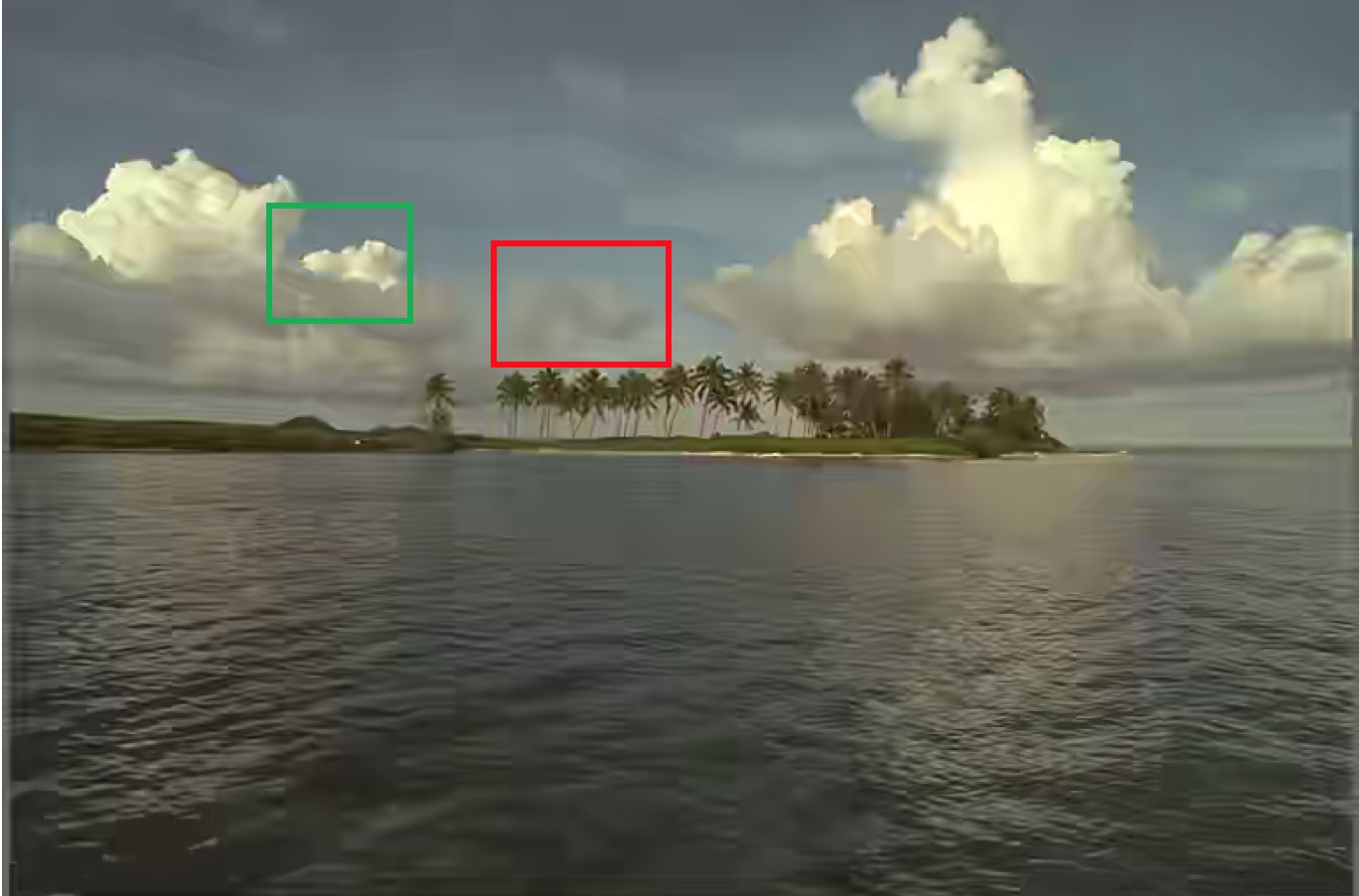} &
   \includegraphics[width=0.20\linewidth, height=20mm]{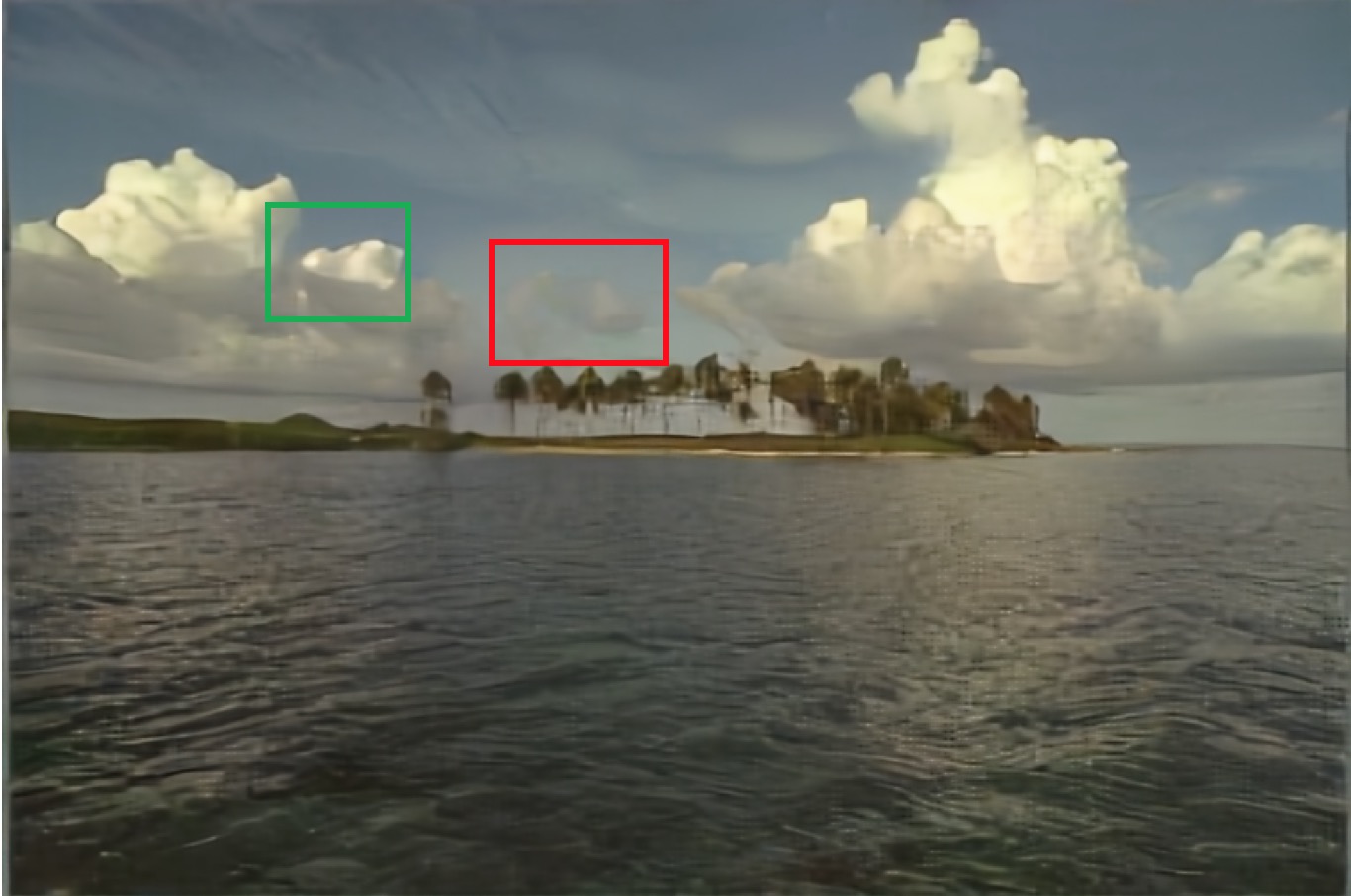} &
    \includegraphics[width=0.20\linewidth, height=20mm]{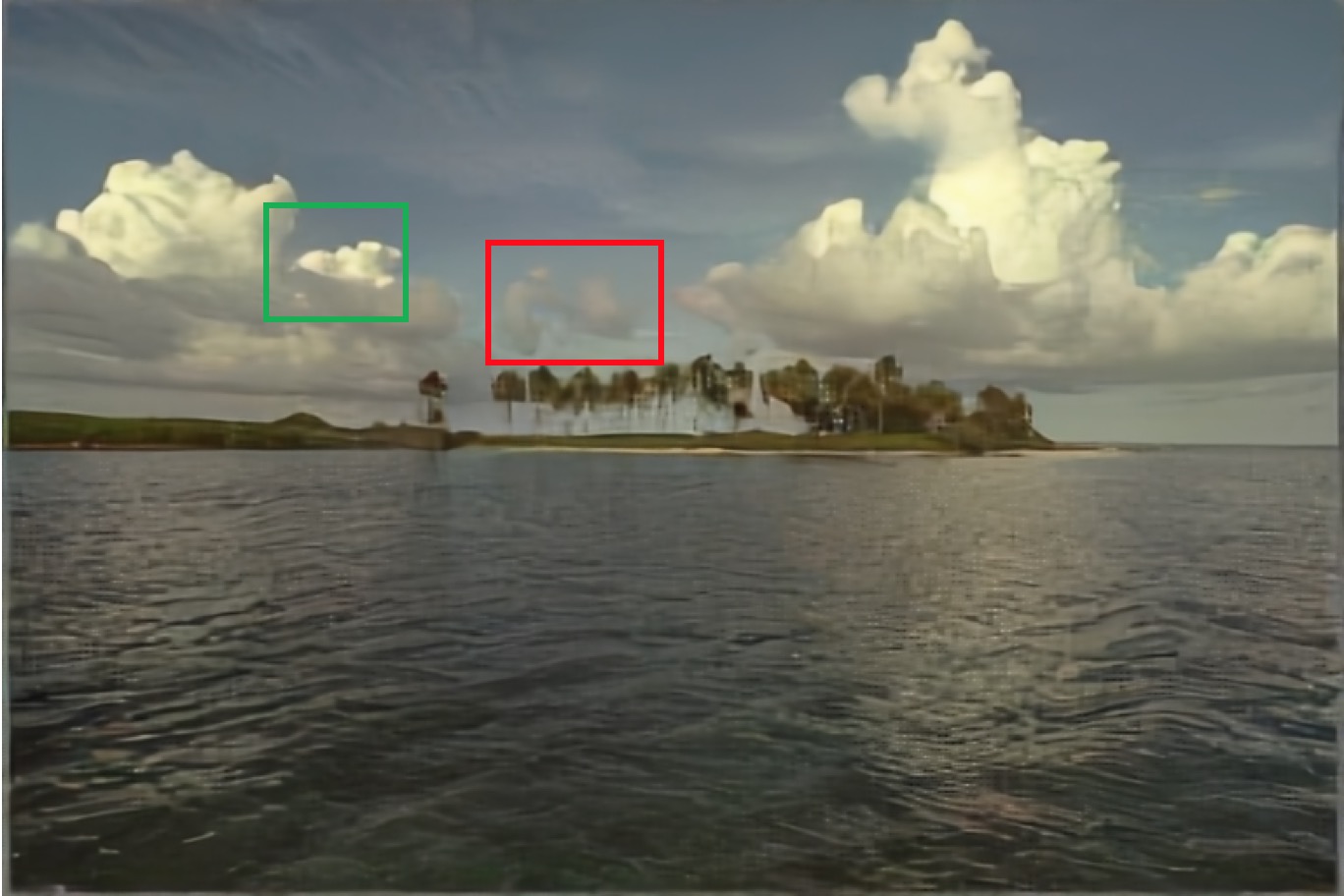}\\
   &
   \fboxsep=0mm
\fboxrule=1.2pt 
  \fcolorbox{green}{black}{\includegraphics[width=0.20\linewidth, height=20mm]{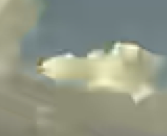}} &
  \fboxsep=0mm
\fboxrule=1.2pt 
 \fcolorbox{green}{black}{\includegraphics[width=0.20\linewidth, height=20mm]{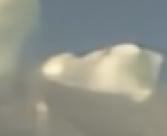}} &
 \fboxsep=0mm
\fboxrule=1.2pt 
  \fcolorbox{green}{black}{\includegraphics[width=0.20\linewidth, height=20mm]{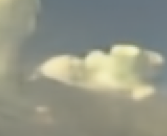}}\\    
   & 
   \fboxsep=0mm
\fboxrule=1.2pt 
  \fcolorbox{red}{black}{\includegraphics[width=0.20\linewidth, height=17mm]{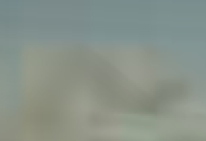}} &
  \fboxsep=0mm
\fboxrule=1.2pt 
  \fcolorbox{red}{black}{\includegraphics[width=0.20\linewidth, height=17mm]{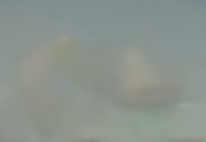}} &
  \fboxsep=0mm
\fboxrule=1.2pt 
  \fcolorbox{red}{black}{\includegraphics[width=0.20\linewidth, height=17mm]{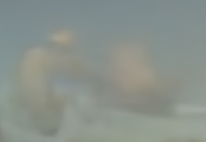}}\\
    & {\small 0.9189/0.112} 	    & {\small 0.9257/0.112} 	    & {\small 0.9323/0.109}  \\
    &  {\small (e) BPG}  			& {\small (f) Ours (SQ)} 		& {\small (g) Ours (TCQ)}  \\
\end{tabular}
\end{center}
   \caption{Qualitative results (MS-SSIM/bpp) from the image \textit{kodim16} on kodak dataset.}
\label{fig:sample}
%\label{fig:onecol}
\end{figure}

\subsection{Comparisons with previous works}
Fig.~\ref{fig:results} shows result comparisons between our approach and other image compression algorithms (Theis \textit{et. al.}~\cite{theis2017lossy}, Ball\'{e} \textit{et. al.}~\cite{balle2016end}, Agustsson \textit{et. al.}~\cite{agustsson2017soft}, Johnston \textit{et. al.}~\cite{johnston2018improved}, Li \textit{et. al.}~\cite{li2018learning}, Mentzer \textit{et. al.}~\cite{mentzer2018conditional}, Cheng \textit{et. al.}~\cite{cheng2019learning}) on two datasets. Despite the simplicity of our network, the results from our model with TCQ show its superior performance at low bit rates. At high bit rates, our results can achieve comparable performance to previous papers except for the latest results in Mentzer \textit{et. al.}~\cite{mentzer2018conditional} and Cheng \textit{et. al.}~\cite{cheng2019learning}. It is probably because at high bit rates, we increase the number of channels of the model, but we do not finetune the training parameters. 

\subsection{Comparisons between TCQ and SQ}
In Tab.~\ref{tab:tcq_scalar_msssim}, we compare the MS-SSIM and PSNR between TCQ and SQ using MS-SSIM loss for training. At the low bit rate (around 0.07 bpp), TCQ can achieve 0.008 in MS-SSIM (0.41dB in PSNR) and 0.005 in MS-SSIM (0.68dB in PSNR) higher than that from SQ on Kodak and Tecnick datasets respectively. We notice that at higher bit rates, the performance gap between TCQ and SQ is less obvious. As the number of channels increases, the learning ability of the model improves as well. The type of quantizer may not be that important for more complex models. 

In Tab.~\ref{tab:tcq_scalar_mse}, we compare the performance between TCQ and SQ using MSE loss as the distortion for training and $ \lambda $ is set to 0.01. A similar trend is observed where TCQ outperforms SQ at the same bit rate.

The pixelCNN++ model used in this paper is not optimal for entropy coding. In~\cite{minnen2018joint}, a context model along with a hyper-network is used to predict $ \mu $ and $ \sigma $ of a set of Gaussian models, which saves more bits than directly using the probability matrix. In our experiment, it gets 0.154 bpp for the model of 8 channels compared to pre-entropy coding with 0.25 bpp on the Kodak dataset. 

\subsection{Qualitative Comparisons}
In Fig.~\ref{fig:sample}, we show results from different codecs. Fig.~\ref{fig:sample} (a) is the original image. In (b), we can clearly see compression artifacts in the JPEG reconstructed image. In (c), (d) and (e), the shape of the cloud is very blurry. For BPG in (e), there are also some block artifacts in the green box sample.
We notice that in (b), (c), (d) and (e), the sky lacks stripped cloud patterns at the upper left corner and there are less ripples in the areas below the trees. Our results in (f) and (g) get generally better perceptual quality. 
%Our method from TCQ achieves the best performance which is 0.9323 in MS-SSIM at 0.109 bpp. 
%In Fig.~\ref{fig:sample1}, (b) and (c) have obvious artifacts in the reconstructed images. (d), (e) and (f) have close performance, however (f) has more clear block patterns on the glasses. Our method from TCQ can achieve best performance which is 0.9742 in MS-SSIM at 0.0871 bpp. 

\section{Conclusion}
In this paper, we incorporate TCQ into an end-to-end deep learning based image compression framework. Experiments show that our model can achieve comparable results to previous works. The comparisons between TCQ and SQ show that TCQ boosts both PSNR and MS-SSIM compared with SQ at low bit rates either using MSE loss or MS-SSIM loss for training.

%\begin{figure}[t]
%\begin{center}
%\begin{tabular}{cc}
%\multicolumn{2}{c}{\epsfig{width=4in,file=Figures/image1}} \\
%\multicolumn{2}{c}{\small{(a)}} \\[1em]
%\epsfig{width=2in,file=Figures/image3.eps} &
%\epsfig{width=2in,file=Figures/image4.eps} \\
%{\small (b)} & {\small (c)}
%\end{tabular}
%\end{center}
%\caption{\label{fig:example}%
%An example figure.}
%\end{figure}

%\begin{table}[tp]
%\begin{center}
%\caption{\label{tab:example}%
%Average PSNR in dB for the ``Coastguard'' video sequence}
%{
%\renewcommand{\baselinestretch}{1}\footnotesize
%\begin{tabular}{|c|c|c|c|c|}
%\cline{2-5}
%\multicolumn{1}{c|}{~}&
%\multicolumn{1}{c|}{2D} &
%\multicolumn{1}{c|}{3D} &
%\multicolumn{2}{c|}{MC-BCS-SPL}\\
%\cline{4-5}
%\multicolumn{1}{c|}{$S_{\text{NK}}$} &
%BCS-SPL & BCS-SPL & $S_{\text{K}}=S_{\text{NK}}$ & $S_{\text{K}}=0.7$\\
%\hline
%0.1 &22.69 &22.76 &23.06 &25.29 \\
%0.2 &24.70 &24.76 &25.78 &27.94 \\
%0.3 &26.37 &26.45 &28.29 &30.15 \\
%0.4 &27.99 &27.95 &30.88 &32.30 \\
%0.5 &29.60 &29.57 &33.58 &34.42 \\
%\hline
%\end{tabular}}
%\end{center}
%\end{table}

\Section{References}
\bibliographystyle{IEEEbib}
\bibliography{egbib}

\end{document}